\documentclass[12pt]{revtex4}
\usepackage{natbib,longtable}
\usepackage{graphicx,array}
\usepackage{delarray}

\usepackage{rotating}

\begin{document}
\renewcommand{\baselinestretch}{1.}
\begin{titlepage}
\title{MASS PREDICTIONS OF ATOMIC NUCLEI IN THE INFINITE NUCLEAR MATTER MODEL$*$}

\author{R. C. NAYAK$^1$ }

\author{L. SATPATHY$^2$}

\medskip
\medskip

\affiliation{1. Department of Physics, Berhampur University, Berhampur-760 007, India, EMAIL: rcnayak00@yahoo.com}
\affiliation{2. Institute of Physics, Bhubaneswar-751 005, India. Email: satpathy@iopb.res.in}
\vskip 5cm

\medskip
\medskip
\begin{abstract}

 We present here the mass excesses, binding energies, one- and two- neutron,
 one- and two- proton and $\alpha$-particle separation energies of 
 6727  nuclei in the ranges
$4\le Z \le 120 $ and $8\le A \le 303 $ calculated in the infinite nuclear 
matter model.
Compared   to our predictions of  1999 mass
table, the present ones  are obtained using larger data base of 2003 mass table of Wapstra and Audi and resorting
to higher accuracy in the solutions of the $\eta$-differential equations of the
 INM model.
The local energy $\eta$'s supposed to carry signature of the characteristic 
properties of nuclei are  found to possess the  predictive capability.
In fact $\eta$-systematics reveal  new magic numbers in the drip-line 
regions giving rise to
new islands of stability  supported by relativistic mean field theoretic 
calculations.
This is a manifestation of a new phenomenon where shell-effect overcomes the 
instability due to repulsive components of the nucleon-nucleon force broadening 
the stability peninsula. The two-neutron separation energy-systematics derived from the present mass predictions reveal a general new feature for the existence 
of islands of inversion in the exotic neutron-rich regions of nuclear 
landscape,  apart from supporting the presently known islands around $^{31}Na$
 and $^{62}Ti$.
The five global parameters representing the properties of infinite nuclear 
matter,
 the surface, the Coulomb and the pairing terms are retained
as per our 1999 mass table.
The root-mean-square deviation of the present mass-fit to 2198  known masses 
is 342 keV, while the mean deviation is   1.3 keV, 
reminiscent of no left-over systematic effects. 
This is a substantive improvement 
over our 1999  mass table   having  rms deviation of 401 keV and mean deviation
of 9 keV for 1884 data nuclei.

\end{abstract}
\maketitle

\vskip 5cm
$*$  This is a briefly enhanced version of the article  just published in Atom. Data and Nucl. Data Tables 98(2012)213-719.
\end{titlepage}
\clearpage


\clearpage

\renewcommand{\baselinestretch}{1.5}

\section{  INTRODUCTION}

Mass formulas always occupy the center-stage in the research arena of nuclear 
physics.
Traditionally mass formulas  have been developed using  the two main
 features of nuclear dynamics, namely, the liquid drop and the
shell features.  The liquid drop has been the mainstay of nuclear physics
since its introduction  by Weizsacher and Bethe  in 1930s.
The liquid -  supposed to represent the nuclear matter composed of the nucleus - 
is a classical one, although in reality it is a quantum liquid consisting of 
interacting many-fermionic system.
At macroscopic level the classical liquid picture has been most useful as a 
reference model.
However it has given rise to some discomfitures, the notable being the 
$r_0-paradox$\cite{hf,jmp} i.e., the discrepancy between the density determined 
by it through the fit to nuclear masses and the actual density 
measured through
electron scattering on heavy nuclei. In the infinite nuclear matter (INM) 
model\cite{ls87,hvh,in88,in99}, the classical liquid drop is
replaced by an INM sphere characterizing the interacting many-fermionic liquid.
Further the model is based on the Extended Hugenholtz-Van Hove 
theorem\cite{hvh,hhv}
of many-body theory, and therefore takes into account the shell
 and the liquid-drop features  nonperturbatively.

The first mass table based on this model was published\cite{in88} in 1988
with predictions of 3481 nuclei. 
The far-off nuclei in the drip-line regions could not be included because of 
accumulated error resulting from  imprecise   solution of $\eta$-equation.
Further development in the INM model was achieved 
 by using better definition of Fermi energies 
leading to  perfect decoupling\cite{prc,prp}   of the finite-size terms  like
surface and Coulomb from  the infinite-nuclear-matter terms 
characterized by   volume and
asymmetry.  This led to the determination\cite{prc,prp} of the saturation properties
of nuclear matter
 i.e., the density, the binding energy per nucleon and the incompressibility
    from  nuclear
masses and thereby resolving\cite{prc,prp} the $r_0-paradox$. Thus the 
basic objective of the INM model was achieved.
 Further for  the successful long-range prediction of nuclear masses, an interactive  network
 extending over the entire nuclear chart was devised\cite{in99} for better 
determination 
of the local energy $\eta$. This  procedure yielded multiple values
of $\eta$ for a given nucleus 
whose ensemble-average was carried out to obtain its unique value.
With these improvements, reliable prediction of nuclear masses up to drip-line 
could be made and
the second mass table containing the data for 7208 nuclei was 
published\cite{in99} in 1999.

Up to 1999 our concern was directed towards the long-standing problem of
the  determination of the  saturation properties of nuclear matter and its 
incompressibility  
 from nuclear masses\cite{prc,prp}.
Once this objective was fulfilled we concentrated on the study of properties 
of  local energy $\eta$, a crucial element in the INM model.
$\eta $ embodies all the characteristic properties of a nucleus, and
therefore carries its finger-print. The $\eta$'s determined in our 1999 mass
predictions were analyzed as a function of neutron number N for a given proton
 number Z revealing\cite{rcn} Gaussian structure for the well-known 
closed-shell nuclei
in the valley of stability and similar pattern for some nuclei in the far-off
 drip-line regions\cite{lsp2}. On the basis of such revelations,
 new neutron magic numbers 100, 150, 164 and proton magic number 78 and 
weakly
 proton magic number  62 and 90 were predicted\cite{lsp2} designating new 
islands of 
stability around $ ^{162}Sm,~ ^{228}Pt$ and $ ^{254}Th$  
 reaffirmed through microscopic study\cite{lsp2} in the Relativistic Mean Field
(RMF) theory and 
Strutinsky 
Shell-Correction calculations. This is suggestive of a new 
phenomenon where shell
stabilizes the instability due to repulsive components (triplet-triplet, 
singlet-singlet) of nucleon-nucleon interaction, 
complementary to the phenomenon of fission isomers and super-heavy elements 
in which  repulsive coulomb instability is overcome by the same.
While the latter elongates the stability peninsula, the former  broadens it.
In another development, comparative study\cite{euro,er2,er3} of the predictive ability 
of different mass formulas carried out with respect to a reference
 mass formula shows random divergences without any common trend. 
This raises the question about predictive ability of mass formulas  and their 
possible use as guidelines for theoretical and experimental
study in unknown regions. Taking the masses calculated in the microscopic study
employing RMF theory as reference masses, it is shown\cite{lsp3} that the divergence
disappears yielding common trend for the prediction of all mass formulas
. Similar trend is also observed\cite{lsp3} when masses predicted in our 1999 mass table
are used as reference. This success  could  be attributed to the very 
structure of the INM model equations.
It is identified that this mass formula is written down in terms
of differential equations. It is a well recognized
fact  that predictive ability is intrinsic in any theory formulated in terms 
of differential equations like that of Newton, Schrodinger and Dirac etc. 
involving rate of change  of physical variables.
This feature of the INM model was not specifically recognized or perceived
so far. We had  implicitly articulated that $\eta$'s of unknown 
regions were obtained by some kind of extrapolation, which was somewhat misleading.

In view of the above developments after 1999, and quite importantly availability of
large number
of new measurement of mass of unknown nuclei in the intervening period, 
prompted us to make a fresh and probably the final mass table. Addition of 314 
newly measured data to the already used 
1884 masses in our 1999 mass table, and more precision  solution of 
$\eta$-differential equation being possible  has resulted 
in the  reliable 
 prediction of masses covering up to drip-line regions with 
rms deviation of
   342 keV and mean deviation of 1.3 keV. This is a substantive improvement 
over our earlier result of 1999 for 1884 mass data  with rms deviation of 401 keV and mean deviation
of 9 keV. 
    
\section{ THE INFINITE  NUCLEAR MATTER MODEL}   


Although the INM model is well-known and described in several papers  
 \cite{ls87,in88,in99,prp}, for easy reference and completeness, a brief account
is presented here.
The INM model is based on quantum-mechanical  infinite nuclear
matter rather than the classical liquid-drop  used in the traditional  
Bethe-Weizsacker (BW) type  mass formulas. 
In this model,  the ground-state energy $E^{\rm
\, F}(A,Z)$ of a nucleus with neutron number $N$, proton number $Z$, mass
number $A$ and  asymmetry $\beta=(N-Z)/(N+Z)$ is considered
equivalent to the energy  of a perfect sphere
made up of infinite nuclear matter at ground-state 
 plus the residual characteristic energy called the local energy $\eta$.
$\eta$ mainly consists of energies due to shell, deformation and diffuseness
etc., which are intrinsic characteristic properties of a given nucleus and can 
be considered as its finger-print. Thus a nucleus possess two categories of 
properties, namely, the global one represented by the INM sphere and 
the individualistic one by $\eta$ (A,Z). So

\begin{equation}
E^F(A,Z) = E^S (A,Z) + \eta (A,Z) 
\label{ifs}
\end{equation}
with
\begin{equation}
E^S(A,Z) = E(A,Z) + f(A,Z) ,
\label{esf}
\end{equation}
where 

\begin{eqnarray}
f(A,Z)& =&
a_s^IA^{2/3}+a_C^I [Z^2-5({3\over {16\pi}})^{2/3}Z^{4/3}]A^{-1/3}
\nonumber \\ & & +a_{\rm ss}^IA^{2/3}\beta^2+
a_{\rm cv}^IA^{1/3}-\delta(A,Z)
\label{faz}
\end{eqnarray}
denotes the finite size effects and $E(A,Z)$ is the energy of the
infinite nuclear matter contained in the sphere. This sphere 
is hereafter referred to as the INM
sphere, and the superscript I stands for the INM nature of the coefficients. 
Here $a_s^I,a_c^I,a_{\rm ss}^I$ and $a_{\rm cv}^I$ are the surface,
Coulomb,  surface-symmetry and curvature coefficients
and 
$\delta(A,Z)$ is the usual pairing term, given by
\begin{eqnarray}
\delta (A,Z) &=& + \Delta A^{-1/2} \quad {\rm for\ even-even\ nuclei}
\nonumber \\
&=& 0 \quad \quad \quad {\rm for\ odd-A\ nuclei} \nonumber \\
&=& - \Delta A^{-1/2} \quad {\rm for\ odd-odd\ nuclei}
\label{delta} 
\end{eqnarray}

Eq. (\ref{ifs} ) now becomes 
\begin{equation}
E^F(A,Z)   =E(A,Z) +f(A,Z) +\eta (A,Z)
\label{inmf}
\end{equation}

Thus the energy of a finite nucleus is written as the sum
of three distinct parts: An infinite part $E$ , a finite part $f$ and a
local part $\eta$ . All these three parts are considered distinct in
the sense that each of them refers to a different
characteristic of the nucleus
and as such, is more or less    independent of each  other.
Eq. (\ref{inmf} ) is our required mass formula which provides a direct
link between finite nuclei to nuclear matter. Its three
functions $E, f$ and $\eta $ are to be determined.

The term $E (A,Z)$  being the property of nuclear matter 
 at the ground state, 
must  satisfy the generalized   \cite{hvh} HVH theorem 
given by 

\begin{equation}
{{E\over A} +\rho {\partial (E/A) \over {\partial \rho }}}
={1\over 2}  [
  (1+\beta)\epsilon_{n}+(1-\beta)\epsilon_{p} ].
\label{hvhg}
   \end{equation}
Using Eq. (\ref {inmf} ),  the INM  Fermi energies $\epsilon _n $ and $ 
\epsilon _p $ 
can be expressed in terms of their counterparts of finite nuclei 
denoted through suffix 'F' as

\begin{eqnarray}
\epsilon_n &=& \epsilon_n^F-{(\partial f/\partial N)}_Z
-{(\partial \eta /\partial N)}_Z\nonumber \\ 
\epsilon_p &=& \epsilon_p^F-{(\partial f/\partial Z)}_N
-{(\partial \eta /\partial Z)}_N
\label{fef} 
\end{eqnarray}
where
$\epsilon_n^F =({\partial E^F /{\partial N}})_Z $
and
$\epsilon_p^F =({\partial E^F /{\partial Z}})_N $ .
Using Eqs. (\ref {inmf} ) and  (\ref {fef} ), the relation 
(\ref {hvhg} ) can be recast as

\begin{eqnarray}
{{E^F\over  A} + {\eta \over  A}} &=& {1\over 2} [
(1+\beta)\epsilon_{n}^F+(1-\beta)\epsilon_{p}^F ] + S(A,Z) \\& &+
{1\over 2}[(1+\beta)({\partial \eta \over {\partial N}})_Z +
(1-\beta)({\partial \eta \over { \partial Z}})_N] ,
\label{fscg}
\end{eqnarray}
where,
 \begin{equation}
S(A,Z)= {f/ A}-{(N/ A)}{(\partial f/ \partial N)}_Z
-{(Z/ A)}{(\partial f/ \partial Z)}_N 
\label{saz}
    \end{equation}
 is a function of all the finite-size coefficients 
$ a_s^I, a_c^I, a_{\rm  ss}^I $ and $a_{\rm cv}^I$ which are global in nature.
 As noted earlier, the local energy  $\eta $ refers to 
a the individualistic  characteristic of the nucleus, while the global function 
$S(A,Z)$
 refers to the bulk properties which are global in nature. Also the above 
Eq. (\ref{fscg})  does
not have any coupled term involving these two functions $\eta$ and $S$.
 Therefore, we make the ansatz of  splitting  the above equation  into the following two equations.

 \begin{equation}
  {E^F\over A} = {1\over 2} [
   (1+\beta)\epsilon_{n}^F+(1-\beta)\epsilon_{p}^F ] + S(A,Z)
    \end{equation}

or equivalently
 \begin{equation}
  S(A,Z) = {E^F\over  A} - {1\over 2} [
   (1+\beta)\epsilon_{n}^F+(1-\beta)\epsilon_{p}^F ] 
\label{fsc}
 \end{equation}
and
\begin{equation}
{\eta(A,Z)\over A}={1\over 2}[(1+\beta)({\partial \eta /
{\partial N}} )_Z +
(1-\beta)({\partial \eta /{ \partial Z}})_N].
\label{etar}
 \end{equation} 
If individually each of these Eqs. (\ref{fsc}, \ref{etar}) is satisfied, then
 the original Eq. (\ref{fscg}) is satisfied.
We would like to stress here that the validity of these two equations 
have been well demonstrated a'posteriori in numerical calculations elsewhere
\cite{ls87,in88,in99}. 

These three equations define the INM model completely;
  Eq. (6)  determines the finite-size coefficients $ a^I_s$ and $ a^I_c  $ etc. 
of the INM
 sphere, Eq. (7) determines the global parameters $ a_v^I $ and $ a_\beta ^I$
  while
Eq. (8) exclusively determines the local energy $\eta$. Thus once the three
functions $E(A,Z), f(A,Z) $ and $  \eta(A,Z)$ are determined, the energy of the 
nucleus (A,Z) is obtained using the mass formula Eq. (3).

\section{ CALCULATION}
It should be emphasized here that the above three Eqs.(6-8) exclusively 
determine the surface, the bulk and the local energies of a given nucleus
 separately. 
The decoupling of these three quantities provides the necessary breakthrough\cite{prc,prp} 
for the correct determination of their respective parameters inhibiting  cross
 correlations amongst them. This led\cite{prc,prp} to the resolution of the $r_0-paradox$
 and determination of the saturation properties including the incompressibility
 of nuclear matter entirely from nuclear masses.
The parameters $ a_{\rm s}^{\,\rm I}, a_{\rm c}^{\,\rm I},
a_{\rm ss}^{\,\rm I}$ and $a_{\rm cv}^{\,\rm I}$ characterizing the INM 
sphere are global in nature.
  The surface symmetry 
 $a_{\rm ss}^{\,\rm I}$ term cancels to the extent of 66\% due to particular
combination of data occurring in it. Further being a 
higher order term,  it is quite  weak compared to the principal term. Combination
of these two features makes it redundant and rather symbolic at numerical level
in the INM model thereby playing no significant role.
We have demonstrated this through exhaustive numerical calculation in references
[\cite{prc,prp}]. 
The same is true for the curvature 
 $a_{\rm cv}^{\,\rm I}$ term.
 So we have dropped   these two terms. We would like to emphasize that in
the INM
model we have adopted the principle to determine all the parameters using
 exclusively the ground-state masses of nuclei only as the model  is intended
 to predict 
the ground-state masses. In the BW-like
 mass models\cite{frdm,frd2} the 
the value of surface-symmetry coefficient $a_{ss}$,
 is usually fixed using properties other than the ground-state masses like 
fission barrier etc. and the curvature term is normally dropped as its coefficient
cannot be determined uniquely. 
 It is true that by dropping the surface-symmetry term in our model,
 isospin-dependence is exclusively taken care of by the bulk symmetry term
only and any  left-over  isospin dependence arising from the surface-symmetry
of a given nucleus is taken care of by the characteristic local energy
  $\eta(N,Z)$.
 Thus this in no way denies any loss of  generality regarding explicit 
absence   of the isospin-dependent surface-symmetry  term  in  
the INM model. 

  We have also made an exhaustive study  to include other higher order 
terms like 
Nolen-Schiffer charge-asymmetry and proton-form factor  etc., however they could not be 
determined uniquely using the masses and the neutron- and proton-separation 
energies in the framework of our model as they exactly cancel out in the INM
Eq. 6(see Ref.\cite{prc,prp} for details). In BW-like mass models the
 coefficients
of these higher order terms are fixed using properties other than the ground-state masses. They vary widely from one model to another. For example,
 in the Finite-Range Droplet (FRDM) model\cite{frdm}, the value of 
charge-asymmetry 
coefficient is 0.436 MeV while in the Finite-Range Liquid-Drop (FRLDM) 
model\cite{frdm}
 its value is 0.10289 MeV. In view of such wide variation  in the value of this
parameter, we feel in general,
  that  it is not desirable to assign some values to them from 
external 
considerations using properties other than the masses. Thus although their 
explicit representation in the INM model is untenable, however their 
collective effect is
implicitly accounted for through the local energy $eta$ being determined  by 
solving Eq. (8) using known experimental masses.
Thus we have altogether four global parameters
$a_{\rm v}^{\,\rm I}$,  $ a_{\rm \beta}^{\,\rm I},~a_{\rm s}^{\,\rm I}$, 
$a_{\rm c}^{\,\rm I}$ and the pairing coefficient $ \Delta $. Since these parameters are universal in 
nature being valid for all nuclei, they have been meticulously determined 
earlier\cite{in99,prc,prp}
 taking  experimental  masses with error bar less than 60 keV to avoid pollution
from other available data with higher error. With these criteria, the masses of
nuclei  in the valley of stability were chosen together with their neutron
and proton separation energies in fitting Eq. (6) and Eq. (7). 
 Although in the present mass table we are using extra 314 nuclei  for mass
predictions compared to
our 1999 mass table, they are not necessarily
 suitable for the determination of the global parameters as they are more exotic in nature. So we have adopted the same values for the global
parameters as determined for the 1999 mass table\cite{in99,prc,prp}. These values are
\begin{eqnarray}
        a_{\rm v}^{\,\rm I}&=&16.108~{\rm MeV}\nonumber\\
        a_{\beta}^{\,\rm I}&=&24.06~{\rm MeV}\nonumber\\
        a_{\rm s}^{\,\rm I}&=&19.27~{\rm MeV}\nonumber\\
        a_{\rm c}^{\,\rm I}&=&0.7593~{\rm MeV}\nonumber\\
        \Delta&=&11.505~{\rm MeV.}\nonumber
\end{eqnarray}
\begin{figure}[!htb]
\includegraphics[width=18cm,height=20cm,angle=270]{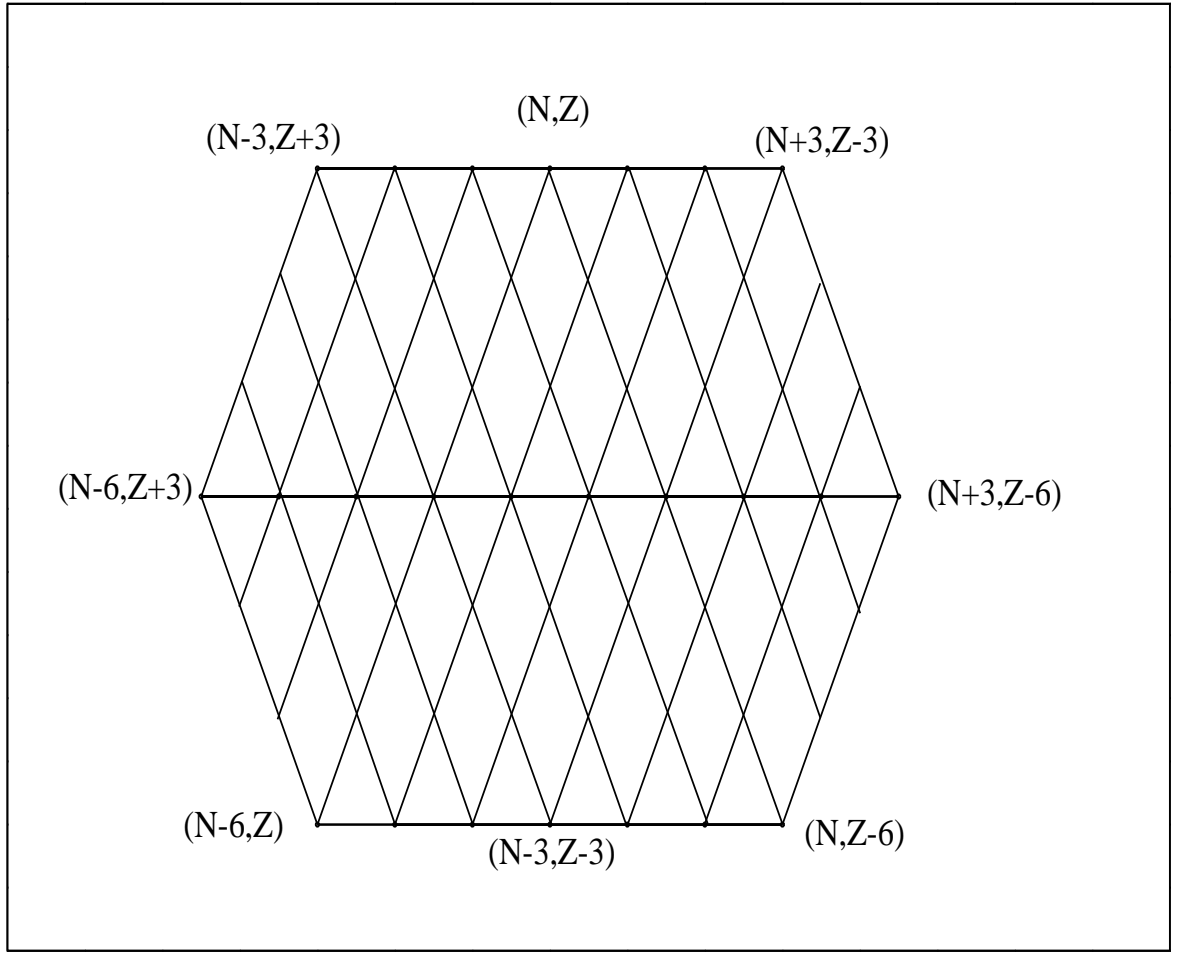}
\caption{ The vertices in the hexagon designated by neutron and proton numbers
 show possible connections among the nuclei taking part in the
 solutions of the local energy $\eta$-equation.
 (\ref{etar}).  }
\label{fig.1}
\end{figure}

Once the global parameters defining $E(A,Z)$ and $f(A,Z)$ 
are known, the empirical values of $\eta$ for all known nuclei are determined
using the mass formula Eq. (\ref{inmf}) .
Prediction of masses of nuclei in the unknown regions amounts to
predicting the $\eta$'s of such nuclei. $\eta$ satisfies Eq. (\ref{etar})
which is a partial differential equation in (N,Z) space. Being expressed
in terms of rate of change of $\eta$ with N and Z, it is endowed
 with good predictive 
ability like well-known equations in Physics. With proper initial conditions
given by experimental values of $\eta$ in known regions, long range 
prediction of $\eta$s can be obtained in unknown regions by solving 
the Eq. (\ref{etar}).
 In all our previous publications
 on the INM model we have presented in detail the construction of the grid in
(N, Z) space and the method of solution, which we do not feel any necessity to 
repeat here. However, we would like to reiterate only the ensemble averaging 
process which was introduced for the 1999 mass table predictions\cite{in99},
 which in fact was crucial for getting proper solution of the $\eta$-Eq. (8)
. Fixing
a hexagonal grid for 58 nuclei in a given region in (N,Z) space (see Fig. 1), 
  Eq. (8) is solved to obtain definite values of
 $\eta$ for those nuclei. Shifting the grid
in all possible directions in (N,Z,A) space, an ensemble of about 70 
alternate values for a given nucleus is generated. The most probable value is
 then obtained
by performing ensemble-averaging using the standard Gaussian weighted method 
given by 

\begin{equation}
\eta = {\sum_i \eta_i~ exp[-\big((\sigma_i-\sigma_0)/\sigma_{rms}\big)^2] 
 \over {\sum_i  exp[-\big((\sigma_i-\sigma_0)/\sigma_{rms}\big)^2]}},
\end{equation}
where $\sigma_0$ is the least of all the $\sigma_i$s for a given nucleus and 
$\sigma_{rms}$ is the rms deviation of all the 2198 known nuclei used in the
 fit. Finally we use the global terms $E(A,Z) $ and $f(A,Z)$ in Eq. (3) 
along with 
the value of $\eta$ so determined to predict the mass of a given nucleus 
in any part of the nuclear chart.

\begin{figure}[!htb]
\includegraphics[width=18cm,height=20cm,angle=270]{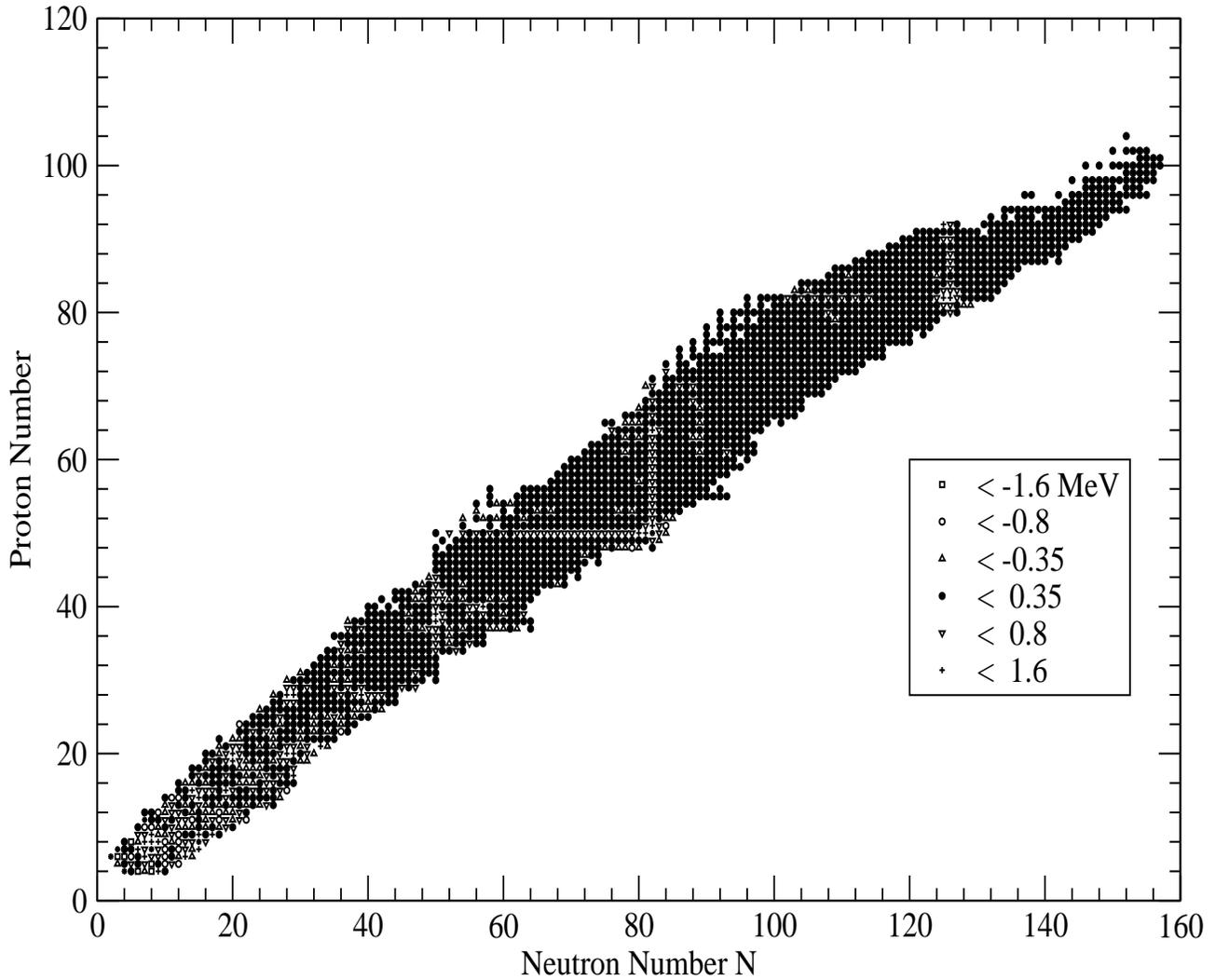}
\caption{ Differences between the experimental (Ref.\cite{aud03}) and
calculated binding energy(BE) for 2198 known nuclei.}
\label{fig.2}
\end{figure}

As before we observe a systematic deviation in the binding energies of N=Z 
nuclei  as a function of mass number as shown in the Fig. 2 of
reference\cite{in99}. There are altogether
31 such nuclei. The binding energy systematics resemble very closely the Wigner contribution 
usually taken into account in the droplet-like models. As before we have taken 
the Wigner-like term given by $ W=W_0/ A \delta_{N,Z}$ as a correction to be 
applied for such nuclei.  We have taken  the value of $W_0$ as 32 MeV as 
determined before\cite{in99}.

Once the global parameters  and the local energies are obtained, the INM mass
Eq. (3) is used to calculate the nuclear masses for the entire nuclear chart.
Thus the present mass table is constructed comprising 6727 nuclei in the range
 $ 4\le Z \le 120$ and  $8\le A \le 303$ extending up to the drip lines.
It may be noted here that our present mass predictions are confined up to
the drip lines and in most cases three to four steps beyond the drip lines in
the nuclear landscape. We feel that this coverage of the nuclear landscape
is sufficient enough for most applications both in Nuclear Physics and Astro-Physics.
That is why the present mass table has 6727 mass predictions compared to 7208
in our 1999 mass table\cite{in99}.
 The rms 
deviation for the fit to  2198 data-nuclei\cite{aud03} is found to be  342 keV
 while  the mean 
deviation is 1.3keV.

\section{ RESULTS AND DISCUSSIONS }

\subsection{ General Features} 

 To check if there is any local fluctuations in the quality of our fit, 
we  have plotted in Fig.2 the  differences between the 
experimental\cite{aud03} 
 binding energies (BE)
 and the fitted ones  termed as BE residuals, in the different
regions of $N-Z$ plane. The various symbols in the figure represent the 
magnitude of the BE  residuals.
 It is satisfying to see that  the magnitude  of the residuals mostly 
lie within 350 keV in accord with  the rms
deviation of  342 keV. In Fig. 3, we have also plotted  the BE
residuals as a function of mass number $A$  to study if any
residual systematics are present, which would signal the deficiencies
of our model. It is interesting to see that the
residuals are  evenly distributed around the A-axis  mostly lying within
350 keV, which is well 
corroborated by the small mean deviation of 1.3 keV.

\begin{figure}[!htb]
\includegraphics[width=18cm,height=20cm,angle=270]{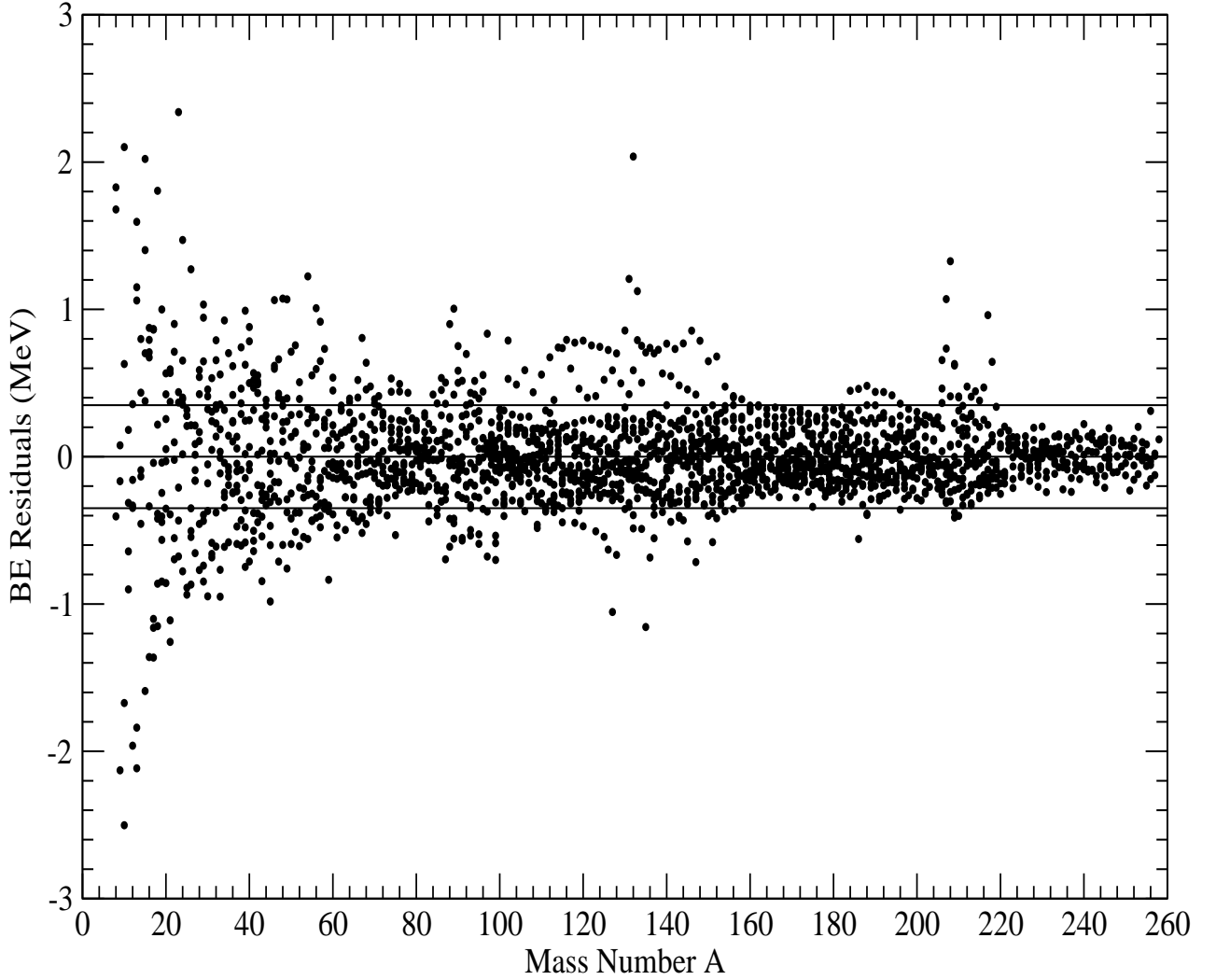}
\caption{  The deviation   as a function of mass number $A$
of the calculated binding energies (BE) from the experimental
data taken from Ref. \cite{aud03}for  2198 nuclei. The two horizontal lines parallel to A-axis drawn at $\pm$0.35 MeV  enclose most of the BE residuals.}
\label{fig:3}
\end{figure}

 We would like to point out below the unique features of this
mass formula which endows it  with such low rms and mean deviation.
\begin{description}

\item {(i)} The INM  model uses not only the nuclear masses like other mass models, 
but also additional two sets of data namely the energy differentials 
$(\partial E/\partial N)_Z$ and $(\partial E/\partial Z)_N$.
 It may be mentioned 
that for any function in general, all
the derivatives at a point are its independent properties apart from
the value of the function
at that point. Thus effectively the model uses three times the mass data 
pertaining to three different properties of nuclei unlike other models for fitting its parameters.

\item {(ii)} The model uses the same set of data two times separately
 in solving Eqs. (6) and (7).
More over in solving Eq. (8) it uses  several tens of  times the same set
of data.
\item {(iii)} Unlike the droplet-like mass formulas, the global parameters
 $a_{\rm v}^{\,\rm I}$ and $a_{\beta }^{\,\rm I}$ are determined by one 
equation (Eq. 7) and the $ a^I_s$ and $ a^I_c  $  etc. are determined by 
another equation (Eq. 6), which allows no cross-correlation among them
thereby leading to
correct values of these parameters.

\item{(iV)}Since HVH theorem is valid for three-body and other multi-body forces,
the INM model being based on it takes them implicitly into account. This is very important
 since three-body forces have been recognized to contribute significantly to nuclear structure  
and the saturation properties. 
\end{description}

In our least-square fit, we have not included the masses of those
nuclei given in the mass Table of Audi and Wapstra\cite{aud03} which are not
experimentally measured but estimated from systematic trends (referred
to as Audi-Systematic Data).  These nuclei mostly lie in the vicinity
of known mass surface with relatively large error bars.
  There are altogether 936  such nuclei. It may be of
interest to compare our predictions with these data as presented in 
Fig. 4. It is satisfying to note that   our predictions compare well with 
the Audi-Systematic Data.

\begin{figure}[!htb]
\includegraphics[width=18cm,height=20cm,angle=270]{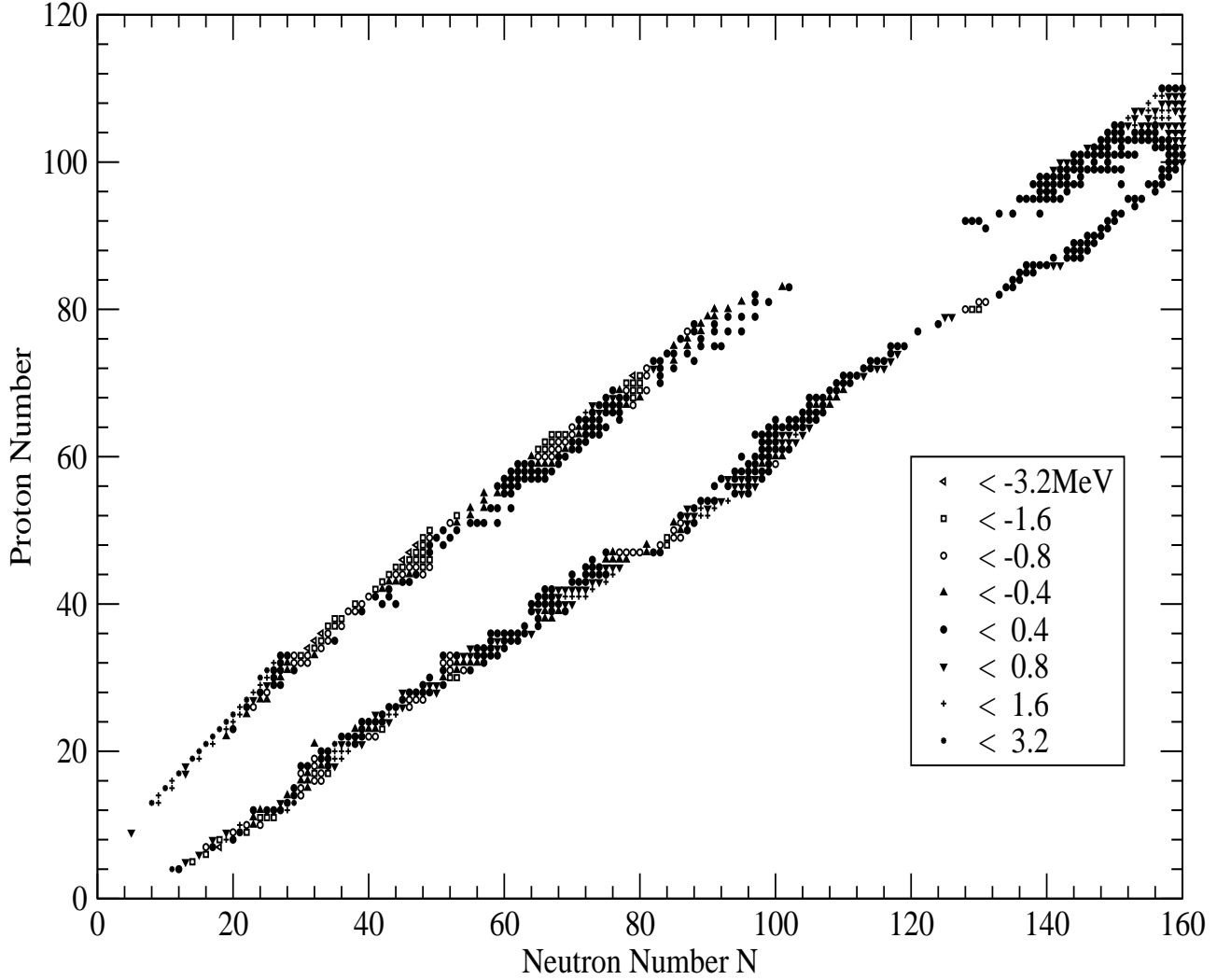}
\caption{ Differences between binding energies (BE) of the Audi-Systematics
 data (Ref.~\cite{aud03})  and the calculated  values of 936 nuclei.}
\label{fig:4}
\end{figure}

\subsection{ Predictive Potential  of the INM model}

	We would like to critically examine  the predictive potential of 
the INM model
basing on two aspects; namely the structural nature of the local energy $\eta$
-systematics, and comparative  global
 analysis of the predictions of different mass models.

 \subsubsection{Structural Nature of $\eta$-systematics:}

 The key element in the INM
 model is the local energy $\eta$ which embodies
 all the characteristic properties of a given nucleus, namely the shell 
effect, the deformation, the diffuseness etc., and possibly other unknown
 properties.     Therefore
 it is expected that study of $\eta$ should reveal  some important features of
 nuclear dynamics.
 In such a study\cite{rcn} the plot of $\eta$(A,Z) as a function of N for a
given Z  shows Gaussian peak at shell-closures in the valley of stability. 
 Such  plots using  the present empirical  values of $\eta$  for all the
 known
 nuclei are presented in  Fig. 5, where
 well-defined Gaussian structures at the magic numbers 50, 82 and 126 
are  prominently seen.
 This feature is suggestive that the $\eta$-systematics in the unknown regions 
may reveal
 new shell-closures in the neutron-rich region close to drip-line which  is
 corroborated in  the following study.

\begin{figure}[!htb]
\includegraphics[width=18cm,height=20cm,angle=270]{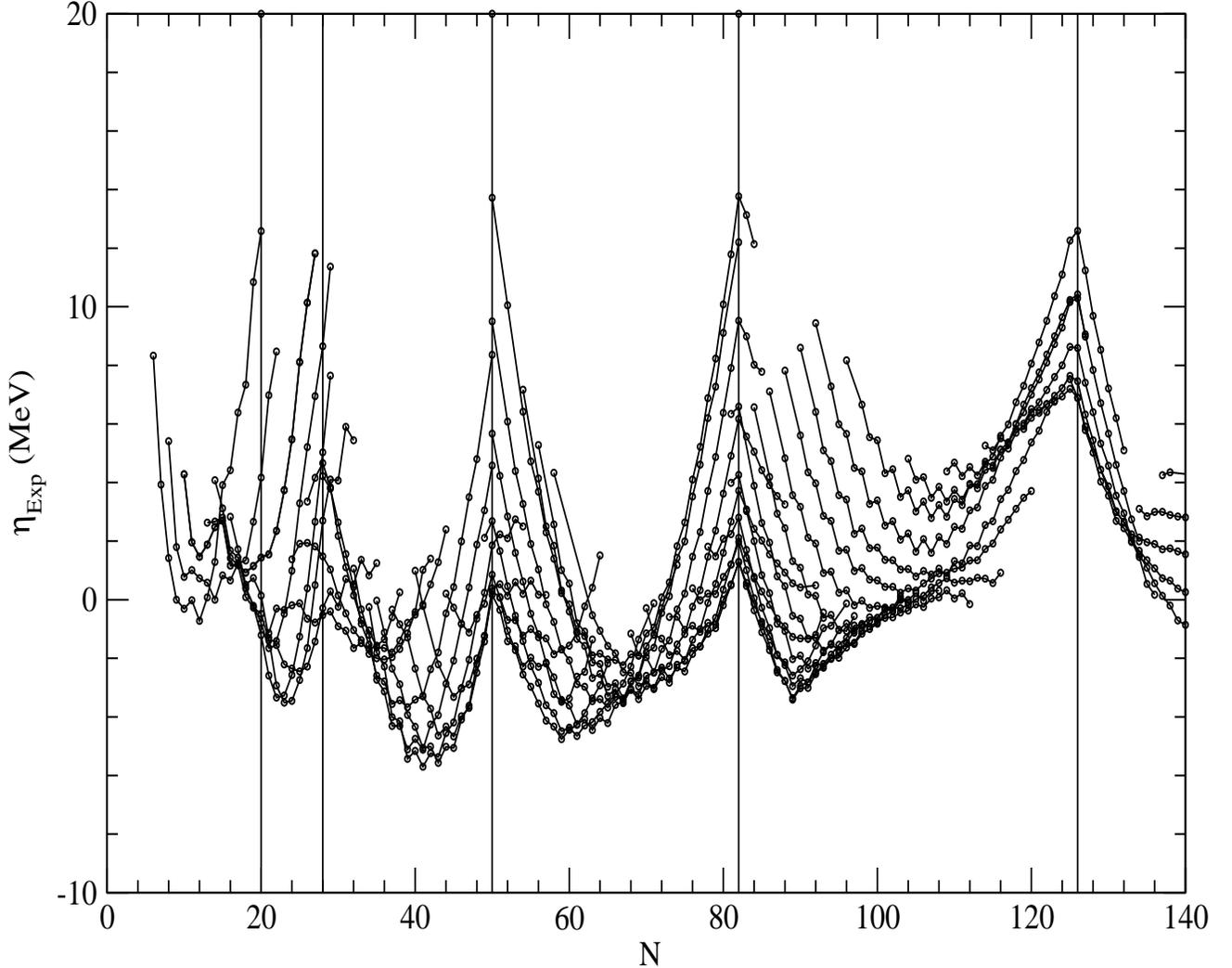}
\caption{ Empirical values of $ \eta $ as  functions of N for all Z showing
structural nature of $\eta$-isolines. Vertical lines drawn at the Magic
Numbers  20, 50, 82 and 126 and the semi-magic number 28 indicate  Gaussian structure.}
\label{fig:5}
\end{figure}
 
In 2004, we\cite{lsp2} have shown the predictive potential  of the INM model by
 predicting
new  neutron magic numbers 100, 150,164 and proton magic number 78 and
 weakly proton-magic 
 numbers 62 and 90 along with new islands of stability around $ ^{162}Sm,
  ^{228}Pt $ and $  ^{254}Th$ in the drip-line regions. This prediction
signals a new phenomenon where shell effect overcomes the instability
 due to repulsive components (triplet-triplet, singlet-singlet) of nucleon-nucleon interaction, which
 is complementary to the phenomena of super heavy elements and fission isomers,  
 in which repulsive coulomb instability is overcome by the same. While the latter
 elongates the stability peninsula the former broadens it.
 This prediction was done on the basis of $S_{2n} $ and $ \eta $ systematics
 obtained in our INM mass predictions. 
 These results  were corroborated through microscopic
 studies in RMF theory and Strutinsky shell correction calculations. Here we
 reinvestigated the same with our new mass predictions. In Fig. 6, we have
 plotted $S_{2n}$ in the upper panel, $\eta$ in the middle panel and the $S_{2n}$-
 differential given by $S_{2n}(N,Z)-S_{2n}(N+2,Z)$ in the lower panel, which
reaffirms our earlier result stated above. 
  Thus the  ability   of the INM model for long range predictions
  looks promising.

\begin{figure}[!htb]
\includegraphics[width=18cm,height=20cm,angle=0]{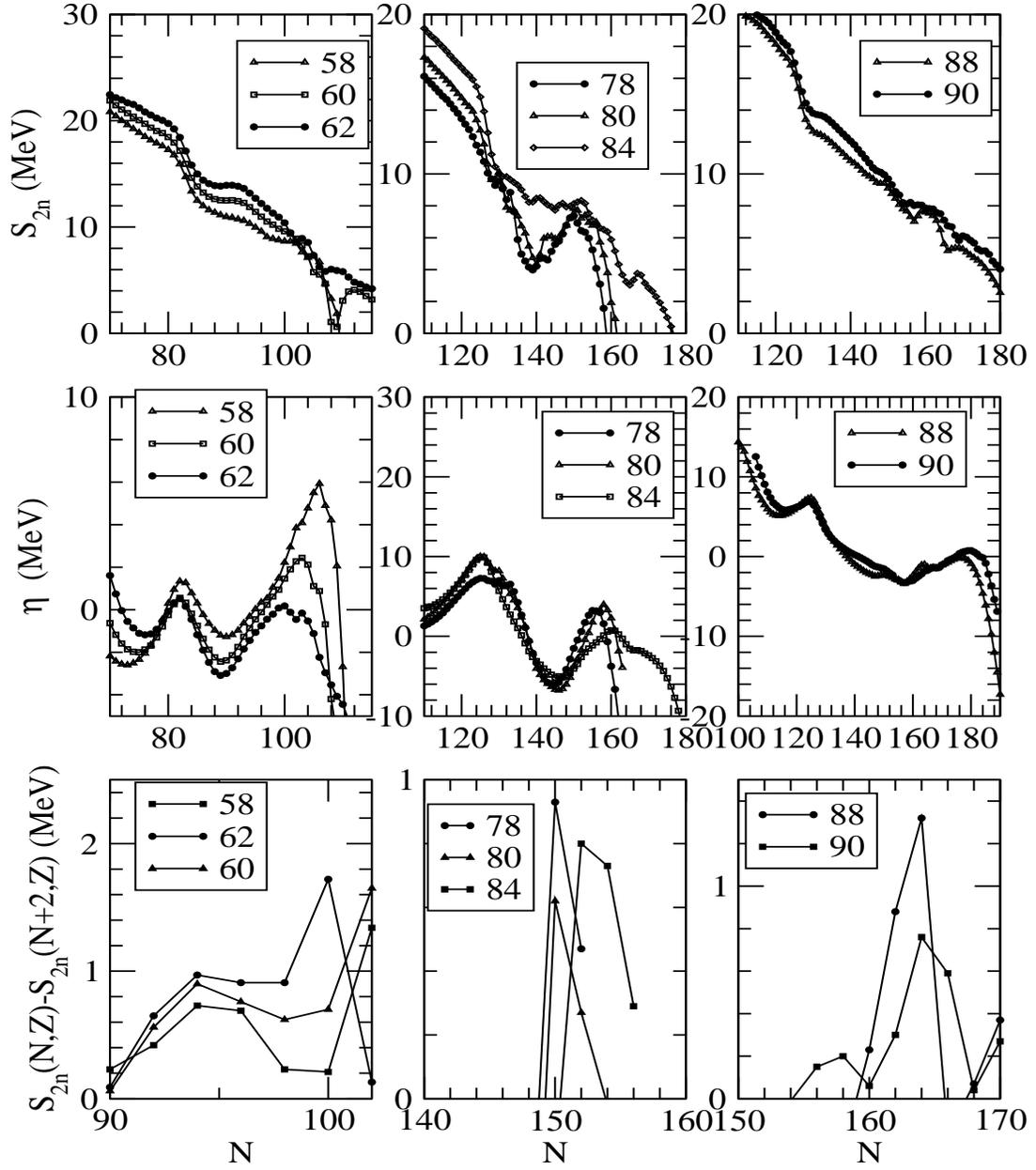}
\caption{ $S_{2n}, \eta $ and $ S_{2n}$-differential (see text ) showing
occurrence of new shell-closures in $^{162}Sm, ^{228}Pt $ and $^{254}Th$. }
\label{fig:6}
\end{figure}

 \subsubsection{Comparative Global Analysis of Mass Models:}
Presently about 2200 nuclei are known and masses of another 5000 to 7000 
nuclei have been
predicted by different mass formulas. Some of them are expected to be 
synthesized in the RIB facilities in the coming years. Therefore the predictive
 ability of different mass formulas is under serious scrutiny\cite{euro,er2,er3}
 in  the recent
years. For this purpose, the predictions of different mass models
\cite{frdm,ten}
for $Sn $ isotopes with neutron number ranging from 45 to 110 have been compared
with those of Duflo-Zucker\cite{dufl} as the reference chosen for its conspicuously low rms
value of 375 keV. 
Such studies\cite{euro,er2,er3} reveal that all mass models\cite{frdm,ten} show good agreement in the known region
in the valley of stability, however the predictions diverge  randomly
without showing any correlation as one moves
away to the unknown regions on either side where experimental results are 
not available. This has raised serious question about the efficacy of mass 
formulas\cite{frdm,ten} as the tools for prediction. For our discussion here we carry out such comparisons more extensively
with respect to Duflo-Zucker, RMF and our present INM model 
predictions not only for Sn isotopes, but also for Pb and Ca isotopes.
Here for a nucleus (N,Z), the BE difference $BED_i (N,Z)$ in the model $i$
 is obtained as
\begin {equation}
BED_i (N,Z)= BE_i (N,Z)-BE_{ref} (N,Z)
\end {equation}
where $BE_i (N,Z)$ and $BE_{ref} (N,Z)$ are the predicted BEs in the 
 mass model $i$ and
 the reference mass model respectively. With Duflo-Zucker  taken as the 
reference model, 
results are plotted in Fig. 7 for  the  Sn, Pb and Ca isotopes separately. 
It can be seen that 
 there is unanimity  of all the mass models\cite{frdm,ten} on good agreement with experiment
in the known region in the valley of stability; however the predictions diverge
 as one moves away to unknown regions on either side. 
Since the masses of the known regions have been used in the fit by all the 
mass models\cite{frdm,ten},
the agreement with data is to be expected. What is worrying is in the unknown
regions where they do not exhibit a common trend like rising and falling, but 
diverge
without any correlation. The divergence with the same intensity is also seen
 when the model of Moeller and Nix et al\cite{frdm} is used as the reference model in Eq.
(10) as is seen in  Fig. 4  of Ref.\cite{euro}.
 This  uncorrelated divergence has raised the question about the efficacy
of mass models\cite{frdm,ten} as a whole\cite{euro,er2,er3}. One is constrained to think whether development
 of different mass models is a parameter game only without  potential for 
reliable predictions!
A probable explanation of this phenomenon may be as follows.

\begin{figure}[!htb]
\includegraphics[width=18cm,height=20cm,angle=270]{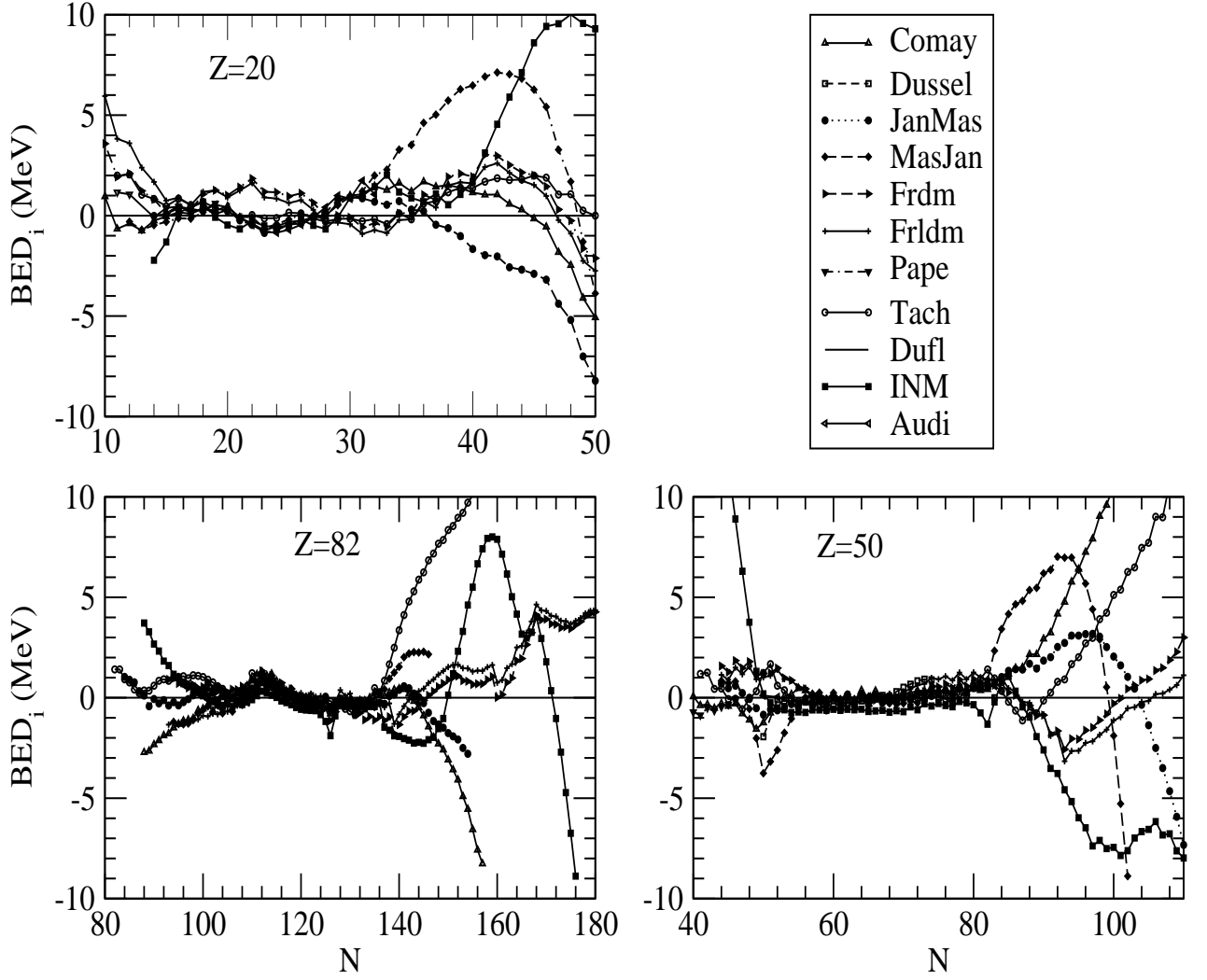}
\caption{ Binding energy difference $BED_i$'s( Eq. 10) in  Ca,Sn and Pb
isotopes of various
Mass Formula\cite{frdm,ten} predictions  with respect to those of
Duflo-Zuker\cite{dufl} as the reference.}
\label{fig:7}
\end{figure}

It is a common feature with most mass models that the degree of success
varies from region to region in (N,Z) space  even in known domains and may be more so in unknown
domains. It is quite likely, the degree of accuracy of  the predictions of 
a given model 
  may not match with the corresponding ones of the reference model in the
same (N,Z) domain resulting 
in randomness with no common trend, which appear as divergences  seen in Fig. 7.

 A microscopic study based on nuclear Hamiltonian may throw light on the above 
issue. We feel masses calculated in Relativistic Mean Field (RMF) theory can 
qualify to serve as a reference mass model. The popular parameter set widely 
used in literature with conspicuous success is the NL3 set\cite{nl3}. The 7 
parameters of 
this Lagrangian have been determined by fitting the data of ten nuclei only,
and with the interaction so determined the predictions\cite{ring} of 1315 even-even nuclei
give rms deviation of 2.6 MeV. It must be noted that the masses of these nuclei
were not fitted unlike other models. So they should be considered as 
predictions
 in the unknown region. Therefore similar validity of this interaction  in
the framework of the RMF for a few hundred nuclei in the true unknown regions
 may be expected in 
 a somewhat lesser scale. The $BED_i$'s calculated with the RMF predictions
as reference for all the isotopes of Ca, Sn and Pb
are shown in Fig. 8. It can be seen that for each chain of these isotopes, 
the predictions in the unknown
regions show a common trend in general for all the mass models\cite{frdm,
ten}, although 
qualitatively they differ from one another. 
 Therefore we may conclude that the predictions 
of mass models in the unknown regions are not necessarily arbitrary or random 
but show definite trends, which therefore  can be used as useful guidelines for theoretical and
 experimental studies. So RMF mass predictions provide a uniform substratum for the
whole (N,Z) domain suitable to serve as a reference.

\begin{figure}[!htb]
\includegraphics[width=18cm,height=20cm,angle=270]{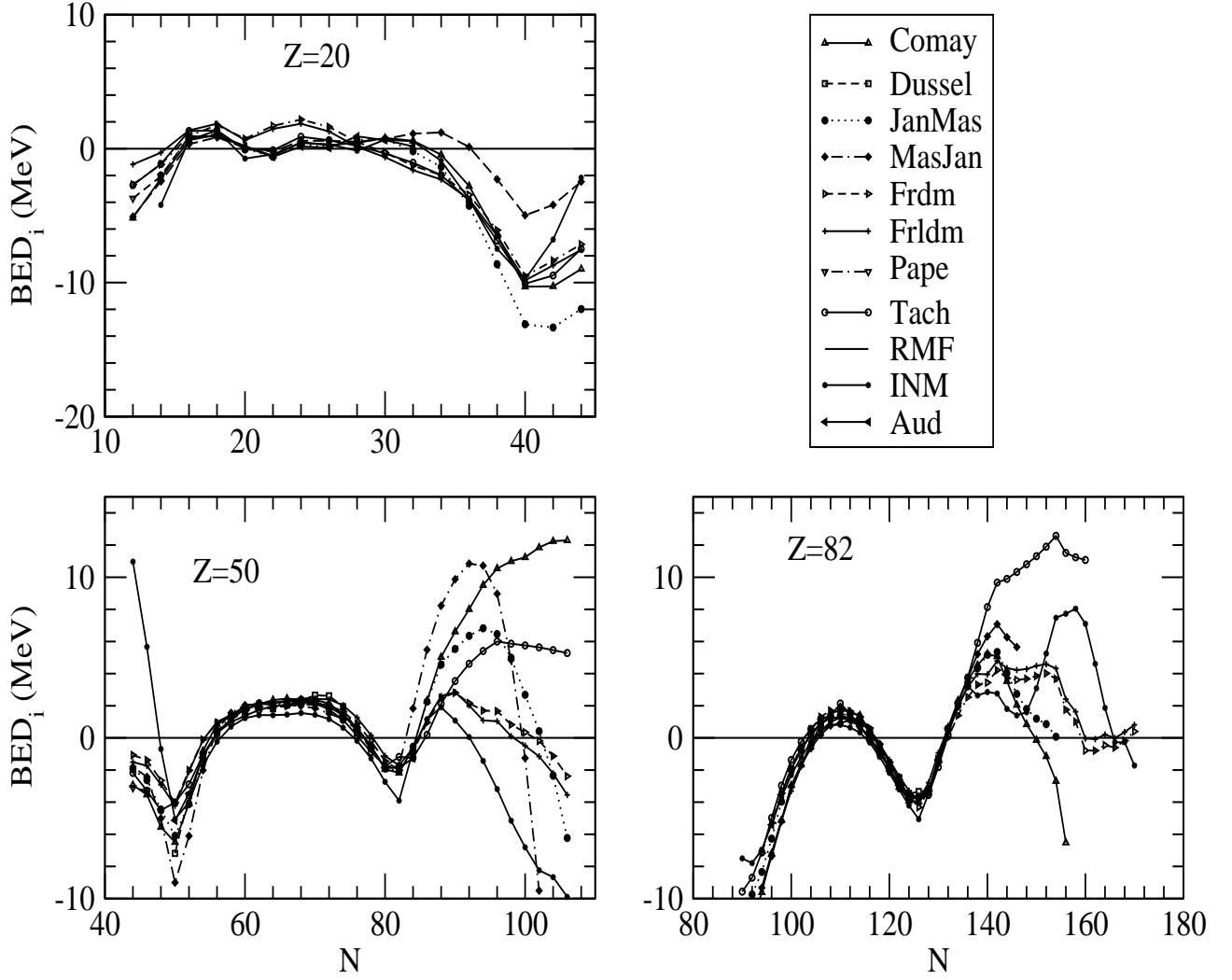}
\caption{ Binding energy difference $BED_i$'s( Eq. 10) in  Ca,Sn and Pb
isotopes of various
Mass Formula\cite{frdm,ten} predictions  with respect to those of
RMF\cite{ring} as the reference.}
\label{fig:8}
\end{figure}

Now in the context of our present mass predictions in the INM model, we use it
 as the reference model in Eq. (10) to
calculate
the  $BED_i$'s  for all those isotopes of Ca, Sn and Pb. The 
results so obtained are
presented in Fig. 9. It is satisfying  to find that no random divergences 
appear in the unknown
regions,  rather  the expected common trends are seen in all the three cases.
 Thus like the RMF mass predictions, INM mass predictions can also be used
as substratum for suitable reference purposes. This implicitly
bears out  our expectation about  the predictive potential  
 of the INM mass model.

\begin{figure}[!htb]
\includegraphics[width=18cm,height=20cm,angle=270]{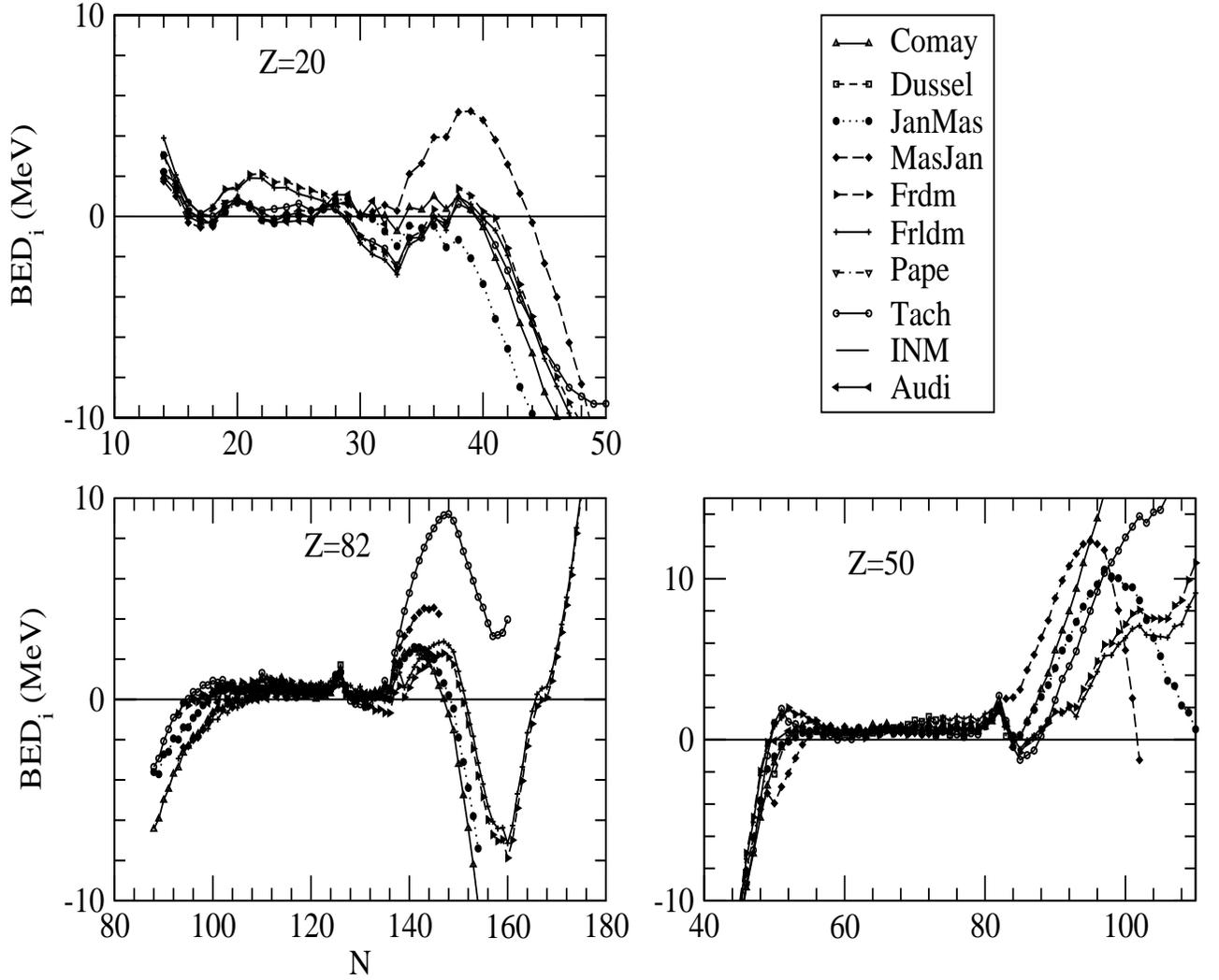}
\caption{  Binding energy difference $BED_i$'s( Eq. 10) in  Ca,Sn and Pb
isotopes of various
Mass Formula\cite{frdm,ten} predictions  with respect to those of
the present  INM predictions as the reference.}
\label{fig:9}
\end{figure}

\subsection{ Shell Quenching in $N = 82$ and $N = 126$ Shells}
\begin{figure}[!htb]
\includegraphics[width=18cm,height=20cm,angle=270]{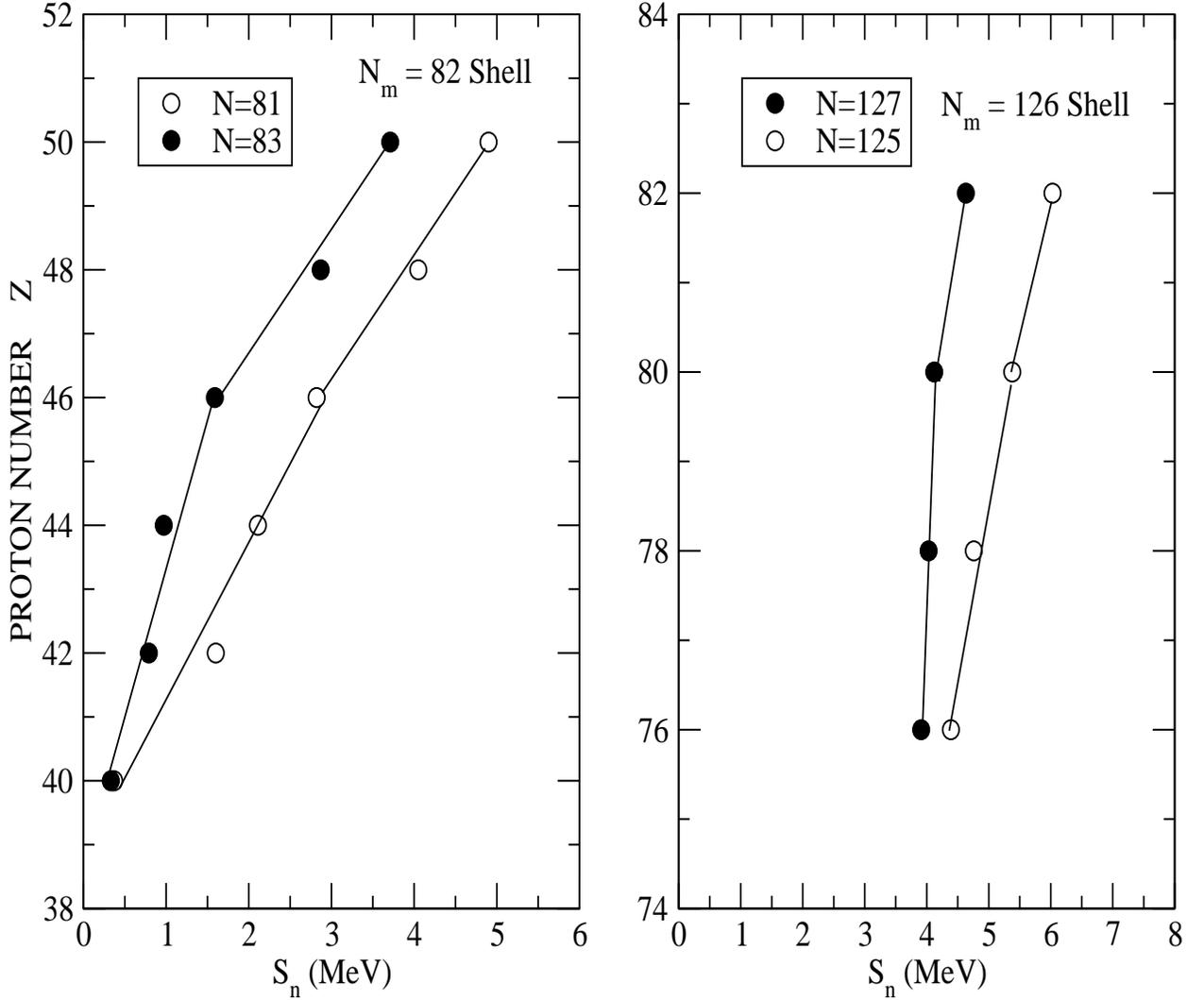}
\caption{ Shell gaps given by $S_n(N_m-1,Z)-S_n(N_m+1,Z)$ for $N_m$=82 and 126
for various proton number Z  showing shell-quenching in N=82 and 126 shells.}
\label{fig:10}
\end{figure}

The shell structure is a very fundamental feature in nuclei. It plays a 
crucial role in building of different models for their description. In the 
valley of stability occurrence of well-defined shell structure 
in the form of magic numbers is well established and extensively studied. In 
the INM model the same  feature
is also seen in the systematics  of local energy $\eta$ when these are plotted as
isolines as shown  in Fig. 5. One can clearly see such sharp-peaked
Gaussian structures at the neutron magic numbers 50, 82 and 126.
 In general, whether such features 
persist in the unknown regions far from stability is of great current interest.
Mass measurements of neutron-rich nuclei in the light and medium heavy regions
have shown\cite{kra} that $N = 20$ shell gap vanishes for nuclei with 
$Z \le 12$. Such
shell quenching has also been seen for the 
$N = 50$ shell\cite{kra2}.  It is hoped that this phenomenon should also be seen for 
 $N = 82$ and $126$ shells. However conclusive proofs presently are lacking 
as  masses of the relevant
nuclei lying  in the drip-line region have not yet been measured
in the laboratory.
 Analysis of the abundances of
various elements in $r$-process nucleosynthesis is suggestive\cite{chn} of the 
  quenching of $N = 82$ shell.
  Our 1999 mass predictions   show\cite{rcn,ls98}  
quenching in $N = 82$, and  $N= 126$ shells.
The present mass predictions reaffirm the same more convincingly.
 In Fig. (10), the shell gaps calculated using the present
mass predictions are shown for $N=82 $ and $ N=126$ shells.
Gradual vanishing of the gaps in both the cases are seen as one moves towards
the drip-line regions.

\subsection{Islands of Inversion}

\begin{figure}[!htb]
\includegraphics[width=18cm,height=20cm,angle=270]{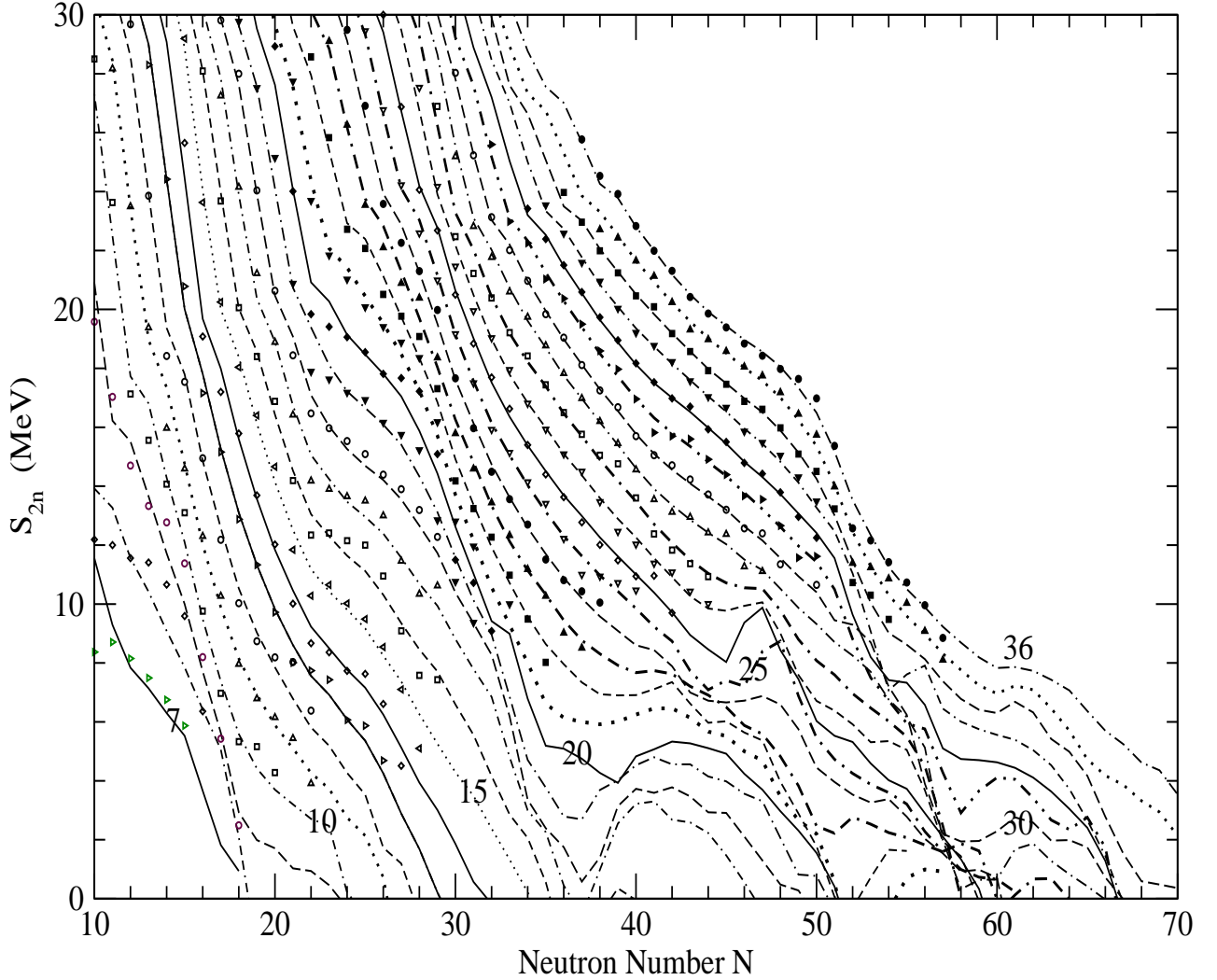}
\caption{ The two-neutron separation energies $S_{2n}$ as function of N  for nuclei with Z= 7 to 36 drawn as solid, broken, dash-dotted etc. lines.
The  corresponding Experimental\cite{aud03}
values where ever available are indicated by various symbols to distinguish the
neighbouring isolines.}
\label{fig:11}
\end{figure}
Neutron-rich nuclei lying in the proximity of the n-drip line are
expected to show some exotic features in nuclear structure with dramatic
departures from the normal trends observed   in the $\beta$-stable valley.
A well-known  phenomenon having such a feature termed as 'island of inversion' 
has been observed\cite{in31} in neutron-rich nuclei around N=20 region 
quite early,  which  manifests in enhanced  binding of those 
nuclei centering around $^{31}Na$.  It has been always a 
challenge to mass formulas and other structure  models to explain this 
phenomenon.
 Extensive theoretical and experimental studies\cite{ext1,ext}
 carried out over
the years have concluded that the N=20 shell-closure in this region is broken
 by the intruder states  from the $pf$-shell,  inducing strong collectivity
giving rise to high deformation.
This has  resulted in enhanced binding for Na and nearby
nuclides in this region.
  Therefore  with the present mass 
predictions in the INM model, we have 
attempted to make a through search throughout the nuclear landscape to see
if our masses can account for the above known island of inversion and also 
identify  possible new such islands. With this view we have plotted the $S_{2n}$
isolines  for all the nuclides from Z=7 to 105 in Figs. 11-13.
 The typical sharp fall of $S_{2n}$
  at the shell-closures clearly reproduce the well-known magic numbers 8, 20, 50, 82 and 126
in conformity with experiment.
However the monotonic decrease with increase of neutron numbers N in the 
$\beta$-stable valley gets arrested
in the neutron-rich  region for Z=10 and 11, agreeing with the observed
island of inversion around $^{31}Na$. Thus our mass predictions
reproduce  this feature reasonably well.
\begin{figure}[!htb]
\includegraphics[width=18cm,height=20cm,angle=270]{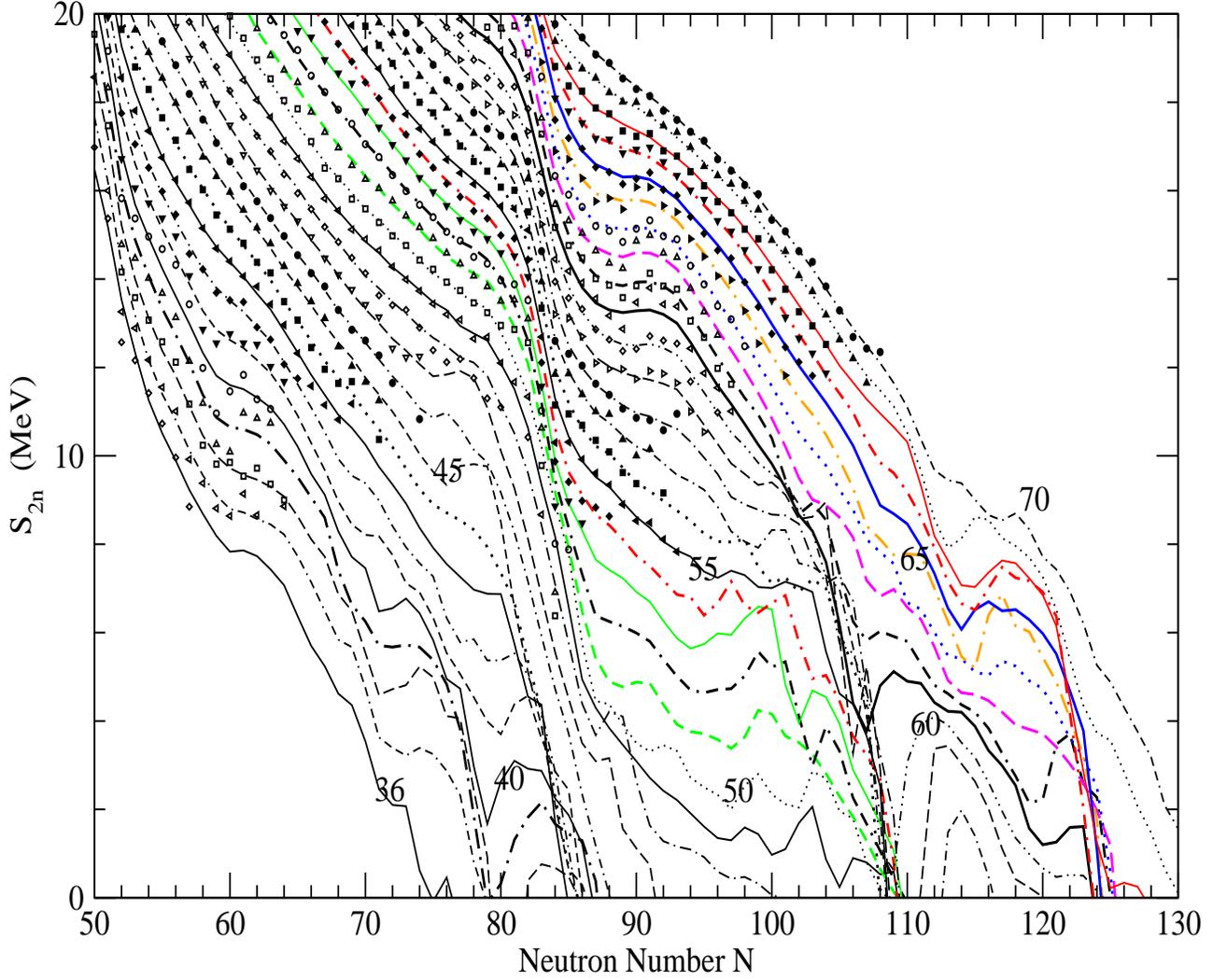}
\caption{ Same as Fig.11 but for Z= 36 to 70.}
\label{fig:12}
\end{figure}

It can also be seen  in   Fig. 11 that there is a region spanned by
 Z=17 to 23 and N=38 to 42,  where 
  $S_{2n}$ isolines exhibit the same feature of enhanced stability
suggesting the existence of another island of inversion.
This  may originate due to 
breaking of N=40 shell by the intruder states from the $sdg$ shell and 
thereby inducing strong deformation.
It is indeed satisfying to note that in the recent years strong
evidences both theoretical and experimental have emerged\cite{in62,inv2}
supporting  the existence 
of this island of inversion centering around $^{62}Ti$.
These two islands of inversion do suggest that these two  may not be the
isolated cases,
and this phenomenon may be a general feature of nuclear dynamics
in the exotic neutron-rich regions close to n-drip line,
 where breaking of shell-
closures  by intruder states   from higher shells are 
highly plausible. In fact
 our extensive $S_{2n}$ systematics in the high-mass region presented
 in Fig. 12 reveal two more 
islands in the heavy-mass region delineated by Z=37-40, N= 70-74; and 
Z=60-64, N= 110-116,
  where these
 may be due to  breaking of N=70 and N=112 shells by the 
intruder states from the $pfh$ and $sdgi$ shells respectively.

\begin{figure}[!htb]
\includegraphics[width=18cm,height=20cm,angle=270]{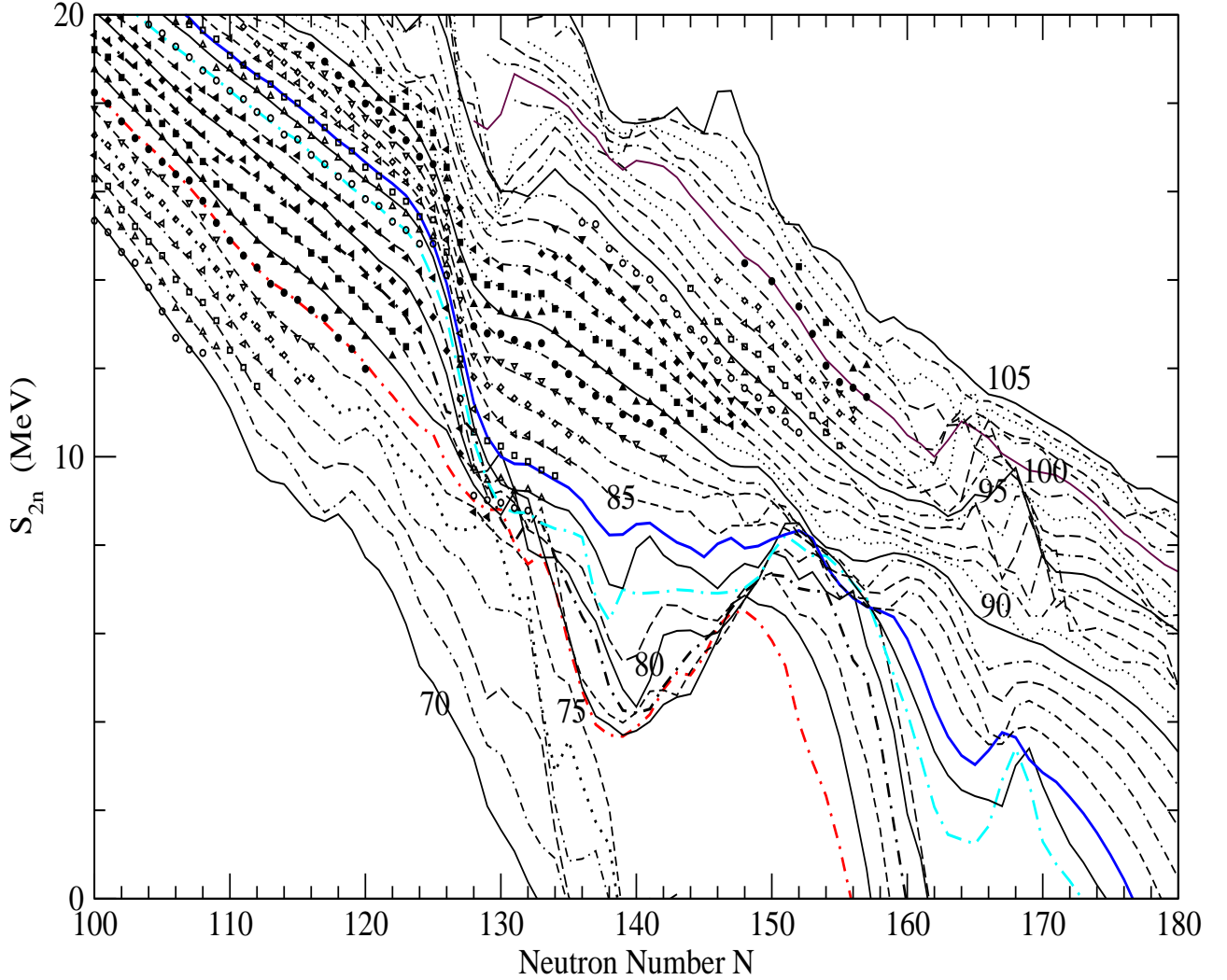}
\caption{ Same as Fig.11 but for Z=  70 to 105.}
\label{fig:13}
\end{figure}

\newpage

\section{ Conclusion  }

The Infinite Nuclear Matter  model of atomic nuclei developed  to its full
potential over the last thirty years has been presented. The model has 
uncovered 
a new facet of the nucleus in which the nuclear matter  characteristic 
has been revealed  to be predominant. The abiding picture since the emergence 
of nuclear physics in 1930s is that the nucleus  can be considered as a 
classical liquid drop  which formed the foundation of Bethe-Weizsaecker(BW)
 mass formula and its more refined version like the droplet model etc.
These liquid drop based models are  traditionalally used to define 
infinite nuclear matter and its saturation properties, determinable 
from nuclear masses. This has lead to  many ambiguities and  inconsistencies
 like
 $r_0$-paradox, indeterminability  of  incompressibility etc.
In contrast,   the INM model 
 has replaced the classical liquid  by a
 quantum mechanical 
many-fermionic liquid  which is the true nature of nuclear matter composed
 of the nucleus, by founding it on the generalized  HVH theorem of many-body 
theory.  Thus 
the  liquid and the shell features, the two main features of nuclear dynamics
could be taken into account non-perturbatively at the microscopic level. 
The model has succeeded in seperating the energy of the nucleus into a 
universal part represented by a sphere made up of infinite nuclear matter
 common to all
 nuclei,
and a charecteristic part $\eta$ called the local energy which varies from nucleus
 to nucleus. The universal part is described  by only four parameters
charecteristsic of the nuclear matter and $\eta$ being determined from nuclear 
masses through the solution of a differential equation of the model.

The effect of three-body  has been included
 inadvertently  by the use of HVH theorem,  which is valid for such forces.
 Because of this proper physics input,
the saturation density of nuclear matter could be obtained from nuclear masses
 in agreement with the value determined from  
electron-scattering   on heavy nuclei, and thereby resolving the long standing
$r_0$-paradox.
The energy per nucleon $ E/A$ of nuclear matter at ground-state has been
 consistently obtained from nuclear masses also. Thus both the  saturation
 properties of nuclear matter $E/A $ and $\rho_{\infty}$ are obtained 
  from the same source using the INM model.
The third property of INM, namely the  Incompressibility could be  determined
in more or less model independent manner using the same mass data that includes 
 nuclear masses,  and neutron and proton Fermi  energies. Thus the data used for
 the determination of all the three saturation properties  numbering about 4000, are well-respected as the most precision 
data in nuclear physics and that too pertaining to ground-state only.

The mass formula based on the INM model has the unique feature that it is 
formulated in terms of differential equations and thereby imbued with the 
ability for prediction of masses in unknown region. The root mean square 
deviation and the mean deviation of the mass-fit to 2198 known masses yields 
respectively 342 keV and 1.3 keV only the lowest in literature. 

The model has introduced in nuclear physics a  new entityi, namely, the local
 energy $\eta$. It contains all the characteristic properties like shell, deformation, 
diffuseness etc and therefore exceeds in scope compared to the usual 
shell-effect 
introduced by Strutinsky. In known regions, it is determined from experimental 
mass, and in unknown regions it is predicted by solution of a differential 
equation using the known ones as boundary conditions. The former being unique
is endowed with the predictive potential, whereas the latter depending 
 on the chosen meanfield is deficient in this respect. The  systematics
of $\eta$ reveals new magic numbers  in exotic neutron-rich nuclei in the dripline 
regions of the nuclear chart. This predictive ability is not contained in the usual
shell effects, the latter being dependent on mean field which varies from interaction to interaction.
 The systematics of $\eta$ predicts three islands of stability 
with N=100, Z=62; N=150, Z=50 and N=164, Z=64 around $ ^{162}SM, ~^{228}Pt $ 
and $ ^{254}Th$ supported through microscopic studies in the framework of RMF
theory and Strutinsky shell-correction calculation. This reveals a new effect
 in nuclear physics, where shell 
overcomes the instability due to repulsive components (triplet-triplet, singlet
-singlet) of nuclear force, analogues to the superheavy elements where repulsive
Coulomb instability is stabilized by the same, while the latter effect elongates
the stability peninsula the former broadens it. 

In comparative global analysis of all the mass models, the unique predictive
 ability of the INM mass model has been established by using its predicted binding energies 
as the reference and studying the systematics of the predicted masses of 
other mass models.

The shell-quenching in N=82 and 126 shells have been found  using the 
INM masses, much anticipated but not predicted by any other mass models  

 The model has also shown the existence of  new islands of inversion delineated by Z=37-40,
N=70-74; Z=60-64, N=110-116; thus showing that the phenomenon is quite general
 and not confined to the well-known islands of inversion around $ ^{31}Na$
for a long time in nuclear physics. Thus in this respect, the INM model has
 been  singulalarly successful.

The unique features of the INM mass model are:

\begin{description}

\item i) It does not use any effective interaction in any way, and thereby it is free
from the vagaries of the uncertainty and  from region to region variation.

\item ii) It uses only the nuclear masses and Fermi  energies determined solely by experiment.

\item iii) It takes into account three-body, and other multibody forces 
(if present) in
 nuclei. 
\item 4) It is built in terms a differential equation and thus equiped with physics for prediction.

\end{description}
The distinctive success of the INM model may be attribited to these features.

\clearpage


\section{EXPLANATION OF TABLE}

\section{ {TABLE}}
	{\bf {Calculated Nuclear Ground-State Binding Energies and Mass 
Excesses, 
          Compared to  Experimental Mass Excesses Where Available along with
One-neutron, Two-neutron, One-proton and Two-proton, $\alpha$-particle
 Separation Energies}}

\begin{center}
\begin{tabular}{ll}
$Z$ & Proton number. The mass table is ordered by increasing proton number.\\
        & The corresponding name of each element is given in parenthesis. \\

$N$	& Neutron number. \\

$A$	& Mass number. \\

BE(MeV)	& Calculated binding energy B(N,Z) of a nucleus (N,Z) using Eq. (3).\\

ME(MeV)	& Calculated mass excess. Nuclei that are unstable against \\
        & one-nucleon and two-nucleon decay are denoted by symbols (\dag) and (\ddag), respectively.\\

ME$_{\rm exp}$(MeV) & Experimental mass excess of Audi-Wapstra\cite{aud03}.
 \\
Err (Mev)	& The  error associated with the Experimental 
	mass excess\cite{aud03}. \\

$S_n (MeV) $ &  Calculated One-neutron  Separation Energies given by  B(N,Z)-B(N-1,Z). \\

$S_{2n}(MeV) $ &  Calculated Two-neutron Separation Energies given by
 B(N,Z)-B(N-2,Z). \\

$S_p $ (MeV) &  Calculated One-proton  Separation Energies given by  B(N,Z)-B(N,Z-1). \\

$S_p $ (MeV) &   Calculated   Two-proton Separation Energies given by
 B(N,Z)-B(N,Z-2). \\

$S_\alpha $ (MeV) & Calculated $\alpha$-Particle Separation Energies  given by
B(N,Z)-B(N-2,Z-2)-28.296. \\

\end{tabular}

\end{center}
\end{document}